\documentclass[12pt,preprint]{aastex}
\usepackage{graphicx,natbib}
\usepackage{amsmath}
\usepackage[usenames,dvips]{color}

\newcommand{\chandra}{\textit{Chandra}}
\newcommand{\lum}{\thinspace\hbox{$\hbox{ergs}\thinspace\hbox{s}^{-1}$}}

\newcommand{\hst}{\textit{HST}}

\begin{document}

\title{Radio Continuum Observations of 47 Tucanae and $\omega$ Centauri: Hints for Intermediate-mass Black Holes?} 

\author{Ting-Ni Lu\altaffilmark{1}, Albert K. H. Kong\altaffilmark{1, 2}} 
\altaffiltext{1}{Institute of Astronomy $\&$ Department of Physics, National Tsing Hua University, Taiwan}
\altaffiltext{2}{Golden Jade Fellow of Kenda Foundation, Taiwan}

\begin{abstract}
We present results of deep radio continuum observations of two galactic globular clusters 47 Tucanae (47 Tuc) and $\omega$ Centauri ($\omega$ Cen) with Australia Telescope Compact Array (ATCA). 
No statistically significant evidence for radio emission was found from the central region for the two clusters. 
However, both clusters show a 2.5$\sigma$ detection near the center that may be confirmed by future deeper radio observations.
The 3$\sigma$ upper limits of the radio observations is 20 and 40 $\mu$Jy/beam for $\omega$ Cen and 47 Tuc, respectively.
By using the fundamental plane of accreting black holes which describes the relationship between radio luminosity, X-ray luminosity and black hole mass, we constrain the mass of a possible intermediate-mass black hole (IMBH) in the globualar clusters. 
We also compare our results with other globular clusters and discuss the existence of IMBHs in globular clusters.

\end{abstract}

\keywords{black hole physics --- globular clusters: individual (47 Tucanae, $\omega$ Centauri)}

\section{Introduction}
Globular clusters are well-known exotic objects factories because of their highly frequent dynamical interactions. 
It has been decades since the prediction was made that intermediate-mass black holes (IMBHs) might exist in globular clusters \citep[e.g.,][]{1970ApJ...160..443W,1975Natur.256...23B,1976MNRAS.176..633F}. 
IMBHs provide a possible connection between stellar-mass black holes and supermassive black holes, with their mass ranging from 10$^{2}$ to 10$^{5}$ M$_{\odot}$.
There have been many conclusive evidences supporting the existences of stellar-mass black holes as in X-ray binaries and supermassive black holes as in active galactic nucleus (AGNs). 
However, the existence of IMBHs in globular clusters is still under debate. 
Although theoretical works provide possibilities of their existences, observational evidences are limited.
There are many formation channels of IMBHs. 
One is that they may be the products of core collapses of population III stars \citep[e.g.,][]{2001ApJ...550..372F}.
IMBHs could also form through runaway merger processes in some young dense globular clusters \citep{2002ApJ...576..899P,2004Natur.428..724P}. 
Moreover, it is possible that in globular clusters, mergers of compact objects, such as stellar-mass black holes through interactions would result in a central black hole with mass $\sim$ 0.001 mass of its host cluster, which falls in the mass regime of IMBHs \citep{2002MNRAS.330..232C}.
As a result, searching for IMBHs in globular clusters is crucial for understanding formation histories of dense stellar clusters, and it may further help us study the stellar dynamics and construct cluster evolution scenarios.
 
There is an established "M-$\sigma$" relation between the supermassive black hole mass and velocity dispersion in the host galaxies based on observations \citep{2002ApJ...574..740T,2005SSRv..116..523F,2009ApJ...698..198G}. 
The correlation implies that the formation of a supermassive black hole is possibly related to the formation of a galaxy. 
If the M-$\sigma$ relation extends to the regime of dense stellar cluster systems, globular clusters may host central IMBHs with masses ranging from 10$^{2}$ to 10$^{3}$ M$_{\odot}$.
One of the formation scenarios of AGNs is that they may be grown from 10$^{2}$ to 10$^{3}$ M$_{\odot}$ formed from direct collapse of population III stars. 
If IMBHs do exist in globular clusters, the M-$\sigma$ relation may link the formation mechanisms of IMBHs and AGNs. 

As far as observations are concerned, detecting IMBHs is achievable through many methods. 
\citet{2009Natur.460...73F} identified an IMBH in the galaxy ESO243-49 by its extremely high X-ray luminosity and variability in the X-ray luminosity and spectra.
For IMBHs in globular clusters, the most common way is measuring velocities dispersion of cluster stars and then applying dynamical a modelling technique to constrain the mass-to-light ratio of globular clusters. 
Another method is detecting possible radio emissions from the center of globular clusters expected from the fundamental plane of black hole \citep{2003MNRAS.345.1057M}.
The dynamical modelling technique can constrain the central "dark" mass in globular cluster, while it fails to distinguish whether the "dark" mass is contributed from a single IMBH or a group of neutron stars, white dwarfs, or stellar-mass black holes.
Moreover, during the past decade, because of the improvement of the instruments, detection of possible radio/X-ray emissions predicted by the black hole fundamental plane becomes possible. 
As a result, several deep radio continuum observations of globular clusters have been carried out to search for IMBHs in globular clusters.

To date, discoveries of IMBHs candidates in some globular clusters have been claimed with different observational techniques (Table 1).
The reported dynamical mass of a possible IMBH in G1 (in M31) is M$_\mathrm{BH}$ = 1.8$\pm$0.5 $\times$ 10$^{4}$ M$_{\odot}$ \citep{2002ApJ...578L..41G,2005ApJ...634.1093G}, while in M15 (in our Milky Way), M$_\mathrm{BH}$ = 1.7$^{+2.7} _{-1.7}$ $\times$ 10$^{3}$ M$_{\odot}$ \citep{2003AJ....125..376G}.  
Nevertheless, the claims of G1 and M15 based on dynamical method have suffered subsequent challenges. 
Theoretical models suggest that it is not necessary to invoke a IMBH to explain the data \citep[]{2003ApJ...582L..21B,2003ApJ...589L..25B}.
Furthermore, latest radio continuum observations on several globular clusters could only set an upper limit on the possible central IMBHs \citep[e.g.,][]{2005MNRAS.356L..17M,2006MNRAS.368L..43D,2008AJ....135..182B,2008MNRAS.389..379M,2010MNRAS.tmp..694C}.
The only radio detection is discovered in G1 \citep{2007ApJ...661L.151U}.
Nevertheless, recent \chandra\ observation of G1 favours the idea that the central object in G1 is an X-ray binary \citep{2010MNRAS.407L..84K}.
Evidences from different detection methods are still not so compelling.
It is thus inconclusive if any IMBH exists in globular clusters.

The two most massive and concentrated Galactic globular clusters, $\omega$ Centauri ($\omega$ Cen) and 47 Tucanae (47 Tuc), have been proposed to harbour a central IMBH.
By using Hubble Space Telescope (\hst) data, \citet{2010ApJ...710.1063V} set a 1$\sigma$ upper limit of 1.2 $\times$ 10$^{4}$ M$_{\odot}$ in $\omega$ Cen, while \citet{2006ApJS..166..249M} put an upper limit of 1000-1500 M$_{\odot}$ on the mass of a possible IMBH in 47 Tuc. 
In addition, previous radio observations set a conservative upper limit of 1000 M$_{\odot}$ and 2060 M$_{\odot}$ for $\omega$ Cen and 47 Tuc respectively \citep{2005MNRAS.356L..17M,2006MNRAS.368L..43D}.
As a consequence, our goal is to set a tighter upper limit on the mass of IMBHs in $\omega$ Cen and 47 Tuc to provide evidences for or against the existence of IMBHs in globular clusters. 

Here we show the results of our deep Australia Telescope Compact Array (ATCA) radio observations of $\omega$ Cen and 47 Tuc. 
We organize our searches of IMBHs in $\omega$ Cen and 47 Tuc in the following sections. 
In $\S2$, we describe the accretion models we adopted to constrain the black hole mass. 
The radio observations, data analysis and results are described in $\S3$. We then constrain the masses of the possible IMBHs in both clusters based on the radio emission upper limits also in $\S3$. 
Finally, we discuss the existence of IMBHs in globular clusters in section $\S4$.

\section{Detecting IMBHs with Radio Continuum Observations}
Recent studies on correlations between X-ray and radio properties of both X-ray binaries and AGNs suggest that the ratio of radio-to-X-ray power increases with black hole mass and decreases with accretion rate \citep[see e.g.,][]{2004A&A...414..895F,2003MNRAS.344...60G,2003MNRAS.345L..19M,2003MNRAS.345.1057M}.
The correlation found in \citet{2003MNRAS.345.1057M} is shown in equation (1), where $\textit{F}_\mathrm{5 GHz}$ is the radio flux at 5 GHz, $\textit{L}_\mathrm{X}$ is the X-ray luminosity, $\textit{M}_\mathrm{BH}$ is the black hole mass, and $\textit{d}$ is the cluster distance.

\begin{eqnarray}
F_\mathrm{5 GHz} &=& 10\left(\frac{L_\mathrm{X}}{3 \times 10^{31}\, \lum}\right)^{0.6} \left(\frac{M_\mathrm{BH}}{100\,\mathrm{M_{\odot}}}\right)^{0.78} 
\nonumber\\
&&\times\left(\frac{d}{10\,\mathrm{kpc}}\right)^{-2}\;(\mu\mathrm{Jy}), 
\nonumber\\
\end{eqnarray}

It is believed that the black holes in globular clusters are likely more massive than stellar-mass black holes, and are accreting at low fractions of the Eddington luminosity (or we would have detect the X-ray emission from the cluster center).
Consequently, the correlation of the fundamental plane may suggest that detection of radio power would be an important indication of IMBHs existing in globular clusters.

Under the circumstances of uncertain X-ray luminosity based on observational constraints, \citet{2004MNRAS.351.1049M} further assumes that the IMBH would accrete intra-cluster gas via a Bondi-Hoyle-Lyttleton (BHL) process \citep{1941MNRAS.101..227H,1944MNRAS.104..273B,1952MNRAS.112..195B}.
As shown by \citet{2003ApJ...587L..35H} and \citet{2003MNRAS.343L..99F}, the accretion efficiency ($\epsilon$) could be as low as 0.1$\%$ of BHL accretion rate, assuming a radiative efficiency ($\eta$) of 10$\%$.
The BHL accretion rate, $\dot{M}_\mathrm{BHL}$, is described in equation (2), where $\textit{n}$ is the gas abundance; $\textit{T}$ is the gas temperature of the globular cluster and assuming a general value of $10^{4}$ K for all globular clusters.
Given the fact that some globular clusters contain a substantial amount of gas measured from pulsars \citep[e.g.,][]{2001ApJ...557L.105F}, one can adopt the gas abundance derived from these globular clusters in equation (2) (see Table 1).

\begin{eqnarray}
\dot{M}_\mathrm{BHL} &=& 3.2 \times 10^{17}\left(\frac{M_\mathrm{BH}}{2000\,\mathrm{M_{\odot}}}\right)^{2}\left(\frac{n}{0.2 \mathrm{\,H\,cm^{-3}}}\right) 
\nonumber\\
&&\times\left(\frac{T}{10^{4}\,\mathrm{K}}\right)^{1.5}\;(\mathrm{g\,s^{-1}}),
\nonumber\\
\end{eqnarray}

Based on the calculation in \citet{2008MNRAS.389..379M,2010MNRAS...M}, the radiative efficiency $\eta$ is expressed as 0.5$\dot{m}$\,c$^{4}$/L$_\mathrm{EDD}$.
This relation is valid for the black holes in the low hard state and the X-ray luminosity $\textit{L}_\mathrm{X}$ $\propto$ $\dot{m}^{2}$.
The $\dot{m}$ here is the mass accretion rate.
Thus, $\textit{L}_\mathrm{X}$ could be estimated by BHL accretion rate $\dot{M}_\mathrm{BHL}$, and expressed with the black hole mass $\textit{M}_\mathrm{BH}$, the gas abundance $\textit{n}$, and the gas temperature $\textit{T}$ of the globular cluster, as shown in equation (3).
If we combine all the above information, the black hole mass could be constrained simply by radio flux and cluster properties.

\begin{eqnarray}
L_\mathrm{X}  = \eta\,\epsilon\,\dot{M}_\mathrm{BHL}\;(\lum).
\end{eqnarray}

$\omega$ Cen was observed previously with ATCA simultaneously at 4.8 and 8.6 GHz over 12 hr, while 47 Tuc was observed briefly (3.5 hr) with ATCA at 1.4 GHz. Both observations failed to detect any central radio source and yielded a 3$\sigma$ rms noise level of 96 $\mu$Jy and 225 $\mu$Jy for $\omega$ Cen and 47 Tuc respectively \citep{2005MNRAS.356L..17M,2006MNRAS.368L..43D}. 
With different combinations of BHL accretion fraction and gas abundance in globular clusters, the radio upper limits correspond to the most probable estimation for the upper limit of the IMBH mass in $\omega$ Cen and 47 Tuc is 2000 M$_{\odot}$ and 1000 M$_{\odot}$, respectively \citep[see][and references therein]{2008MNRAS.389..379M,2010MNRAS...M}. The results could not exclude the possibility that there is no IMBH in $\omega$ Cen or 47 Tuc, which motivated us to propose deeper radio observations on both targets.

\section{Radio Observations}
\subsection{Data Analysis}
47 Tuc and $\omega$ Cen were observed by ATCA during 24 to 25 January 2010. 
The observations were performed simultaneously at frequencies 5.5 GHz and 9 GHz with configuration 6A (with the baseline ranging from 337 m to 6 km) and the upgraded Compact Array Broadband Backend (CABB). 
The data was taken with the CFB 1M-0.5k correlator configuration with 2 GHz bandwidth and 2048 channels, each with 1 MHz resolution.
The primary calibrator used was 1934-638 for both globular clusters, while the phase calibrators were 2353-686 and 1320-446 for 47 Tuc and $\omega$ Cen, respectively.
At the start of each observation, we observed 1934-638 for 10 min. The phase calibrator was observed every 15 min.
We use \texttt{MIRIAD} \citep*{1995ASPC...77..433S} to analyse the data with standard processes.
 When loading the ATCA data into \texttt{MIRIAD}, we use \texttt{atlod} with options \texttt{birdie}, \texttt{xycorr}, \texttt{rfiflag}, and \texttt{noauto}, which represents flagging out the channels affected by the ATCA self-interference, correcting the phase difference between the X and Y channels, discarding any autocorrelation data, and automatically flagging out frequency bands that are known to be heavily affected by RFI.
We then perform the standard data reduction steps, including bandpass, phase and amplitude calibrations.
When producing the dirty maps, we use multi-frequency synthesis (MFS) method \citep*{1999ASPC..180..419S} and natural weighting to suppress the noise.
The effective on-source integration time of 47 Tuc is 11 and 9 hours, while the effective on-source integration time of $\omega$ Cen is 18 and 16 hours for frequencies 5.5 GHz and 9 GHz, respectively. 
The field of view of ATCA with configuration 6A is $\sim$ 10$^{\prime}$ and 5$^{\prime}$ for 5.5 GHz and 9 GHz, respectively, which both cover the whole area within the half-mass radius of 47 Tuc (r$_{h}$ = 3$\farcm$17) and $\omega$ Cen (r$_{h}$ = 5$\farcm$00). 
The spatial resolution of the radio observations can reach $\sim$ 1$^{\prime\prime}$ to 2$^{\prime\prime}$. 

\subsection{Results}
Besides background quasars, we did not detect any radio sources at or near the central region of $\omega$ Cen and 47 Tuc with radio emissions higher than 3$\sigma$. 
In naturally weighted maps, the rms noise level for $\omega$ Cen is 7 (5.5 GHz) and 11 (9 GHz) $\mu$Jy/beam; while for 47 Tuc, it is 17 (5.5 GHz) and 20 (9 GHz) $\mu$Jy/beam. 
We combined the observations with different frequencies in order to obtain a lower rms noise level at frequency 6.8 GHz. 
As a result, the observations reached a rms noise level of 6.5 and 13.3 $\mu$Jy/beam, giving a 3$\sigma$ upper limit of 20 and 40 $\mu$Jy/beam for $\omega$ Cen and 47 Tuc, respectively. 
The brightest millisecond pulsar in 47 Tuc has a radio flux of $\sim$ 370 $\mu$Jy at 1.4 GHz \citep{2004MNRAS.348.1409M}.
Assuming a spectral index $\alpha$ = -1.8 for pulsar spectrum \citep{1998ApJ...501..270K}, it would have a flux of $\sim$ 21 $\mu$Jy at 6.8 GHz. 
Thus the millisecond pulsars in 47 Tuc are all under the detection threshold of our observation.
We overlapped the positions of the X-ray sources detected in $\omega$ Cen \citep{2009ApJ...697..224H}, and there is no any radio source with their signal-to noise ratio higher than 3 matched to the 180 \chandra\ X-ray sources.

Figure 1 shows the \hst\ optical images overlaid with ATCA radio map contours of 47 Tuc and $\omega$ Cen. 
The radio position of 47 Tuc and $\omega$ Cen has an estimated error of $\sim$ 1.4$^{\prime\prime}$ by 0.6$^{\prime\prime}$ and 1.1$^{\prime\prime}$ by 0.7$^{\prime\prime}$ respectively, derived by dividing the beam size by the signal-to-noise ratio.
Although we did not find any statistically significant radio emission, there is a 2.5$\sigma$ detection near the center of both globular clusters ($\sim$ 0.4$^{\prime\prime}$ from 47 Tuc center and $\sim$ 0.7$^{\prime\prime}$ from $\omega$ Cen center). 
We calculated the average source number within the error ellipse of the globular cluster center by computing the area ratios of the error ellipse to the field of the 10$^{\prime\prime}$ $\times$ 10$^{\prime\prime}$ region and also the total number of the sources (with their signal-to-noise ratio larger than 2.5) inside the field of view of the 10$^{\prime\prime}$ $\times$ 10$^{\prime\prime}$ region. 
With the assumption of Poisson distribution, we obtained the probabilities of finding one or more radio sources inside the error circle, which is $\sim$ 2$\%$ for both 47 Tuc and $\omega$ Cen. 
That is to say, the probability of chance coincidence is quite low and the nearby source could really be the central radio source.

\subsection{Constraints on IMBHs mass}
The 3$\sigma$ upper limits of radio emissions could be applied to the mass constraints on IMBHs in globular clusters, using the fundamental plane of black hole activity described in section 2.
We estimate the most probable mass and conservative mass of the IMBHs in $\omega$ Cen and 47 Tuc as described in \citet{2008MNRAS.389..379M}, with the accretion efficiency $\epsilon$ = 0.1$\%$ -- 3$\%$ and the radiative efficiency described in \citet{2008MNRAS.389..379M}.
The gas abunance in globular clusters could be estimated either based on dispersion measurements of pulsars, like M15 and 47 Tuc \citep{2001ApJ...557L.105F}, or based on the stellar mass loss rate \citep{2001ApJ...550..172P}.
We adopt the gas abundance calculated from \citet{2001ApJ...550..172P} in our calculation for $\omega$ Cen, while for 47 Tuc, we use the gas abundance from pulsar measurements as their conservative estimation.
The resulting mass of IMBH is 1100--5200 M$_{\odot}$ and 520--4900 M$_{\odot}$ for $\omega$ Cen and 47 Tuc, respectively.
We list the summary of recently observed globular clusters and our observation results in table 1.

\section{Discussion}

To date, several radio continuum observations have been performed on a few globular clusters in order to detect possible IMBHs in cluster center. 
There is a radio source detected in G1.
However, it is still under debate that whether the radio source in G1 is related to an IMBH or a LMXB \citep{2010MNRAS.407L..84K}.
Except for G1, no central radio source has been detected in the other globular clusters.
Nevertheless, the non-detections did not rule out the existence of IMBHs in globular clusters.
The radio flux upper limits are used to constrain the IMBHs masses, regarding certain assumptions of gas properties in globular clusters and accretion model.
Based on \citet{2001ApJ...550..172P}, the gas abundance in globular clusters could vary from 0.1 to 1 H\,cm$^{-3}$.
The accretion efficiency of the putative IMBH in G1 is just below 1$\%$ \citep{2007ApJ...661L.151U}.
Moreover, according to the standard accretion disk theory, the IMBH in globular clusters would be in an extremely low state, while as suggested in \citet{2010MNRAS.tmp..694C}, both the accretion efficiency and radiative efficiency could vary in a wide range.
The scatter in the black hole fundamental plane correlation would also contribute errors in IMBHs masses estimate.
Combined all these uncertainties, the estimation of IMBH masses is actually model-dependent and under strong model assumptions.

Despite of the large uncertainties in the estimation of IMBHs masses based on radio observations, there are discrepancies between the IMBHs masses estimated from accretion model constraints and dynamically inferred IMBHs masses, especially for the case of $\omega$ Cen (see table 1). 
The measurement of IMBHs masses based on stellar dynamics is limited to the instrument ability, which could not achieve high enough spatial resolution, although it provides a more accurate estimation on the IMBHs masses. 
We further compare the upper limits of IMBHs masses of some globular clusters to the M-$\sigma$ relation of galaxies (figure 2).
The upper limit for the mass of the IMBH in 47 Tuc is consistent with the M-$\sigma$ relation of \citet{2009ApJ...698..198G}, while for $\omega$ Cen, the upper limit of the IMBH mass is within the region of the M-$\sigma$ relation of \citet{2005SSRv..116..523F}.
As the true masses of IMBHs must be lower than the upper limits, $\omega$ Cen, 47 Tuc and M15 do not show inconsistency of the M-$\sigma$ relation.
There is no reason that globular clusters should obey the same relation which is confirmed to exist in massive galaxies.
However, the consistency may suggest similar formation mechanisms or contents between globular clusters and galaxies.
If we could lower the upper limits of the IMBH masses in globular clusters, it may provide evidences of different formation mechanisms between globular clusters and galaxies.

In summary, we estimate the 3$\sigma$ upper limit for the mass of the central IMBH in $\omega$ Cen of 1100--5200 M$_{\odot}$, and for 47 Tuc, we estimate the 3$\sigma$ upper limit of 520--4900 M$_{\odot}$. 
The estimations strongly depend on the accretion model and the gas properties of globular clusters.
We detect a 2.5$\sigma$ radio emission near the center of both globular clusters. 
Future radio observations with a sensitivity improvement to $\sim$ 5 $\mu$Jy may be able to confirm if the sources are real or further constrain the parameters of accretion model.

\begin{acknowledgements}
This project is supported by the National Science Council of the Republic of China (Taiwan) through grant NSC 96-2112-M-007-037-MY3 and NSC 99-2112-M-007-004-MY3.
The Australia Telescope is funded by the Commonwealth of Australia for operation as a National Facility managed by CSIRO.
\end{acknowledgements}

\begin{table}
\centering{\footnotesize
\caption{Recent Radio continuum observations on globular clusters}
\begin{tabular}{lccccccc}
\hline
\hline
Cluster & Distance & $n_\mathrm{H}$ & $T_{gas}$ & $F_\mathrm{R, 5 GHz}$ & $M_\mathrm{BH, rad}$ & $M_\mathrm{BH, dyn}$   \\
 name & (kpc) & (H\,cm$^{-3}$) & (Kelvin) & ($\mu$Jy) & ($M_{\odot}$) & ($M_{\odot}$) \\
\hline
$\omega$ Cen &5.3&0.044&$10^{4}$&20&5200/1100&12000 \\
47 Tuc &4.5&0.28/0.07&$10^{4}$&40&4900/520&1500 \\
NGC 6388 &10.0&0.1&$10^{4}$&81&1500/735&5700 \\
NGC 2808 &9.5&0.26&$10^{4}$&162&8500/1800&2700 \\
M15 &10.3&0.42/0.2&$10^{4}$&25&4900/700&1000 \\
M62 &6.9&0.41&$10^{4}$&36&2900/600&3000 \\
M80 &10.0&0.21&$10^{4}$&36&5300/1100&1600 \\
NGC 6397 &2.7&0.16&$10^{4}$&216&4300/900&50 \\
G1 &780&$\sim$1&$10^{4}$&28&4500&18000 \\
\hline
\hline
\end{tabular}
}
\par
\medskip
\begin{minipage}{0.95\linewidth}
NOTES. --- 
The radio flux $F_\mathrm{R, 5 GHz}$ is the 3$\sigma$ upper limit except for G1, which is a detection.
The radio flux for NGC 6388 is taken from \citet{2010MNRAS.tmp..694C}, for NGC 2808 from \citet{2008MNRAS.389..379M}; for M 15, M 60 and NGC 6266 from \citet{2008AJ....135..182B}; for NGC 6397 from \citet{2006MNRAS.368L..43D}; for G1 from \citet{2007ApJ...661L.151U}.
The dynamical black hole mass $M_\mathrm{BH, dyn}$ adopted for $\omega$ Cen is from \citet{2010ApJ...710.1063V}; for 47 Tuc from \citet{2006ApJS..166..249M}; for NGC 6388 from \citet{2007ApJ...668L.139L}; for M15, M62 and M80 come from \citet{2008AJ....135..182B}; for NGC 2808 and NGC 6397 from M-$\sigma$ relation of \citet{2002ApJ...574..740T}; for G1 from \citet{2002ApJ...578L..41G}.
The radio black hole mass $M_\mathrm{BH, rad}$ for NGC 6388 is from \citet{2010MNRAS.tmp..694C}.
\end{minipage}
\end{table}

\begin{figure}
\includegraphics[scale=1.46]{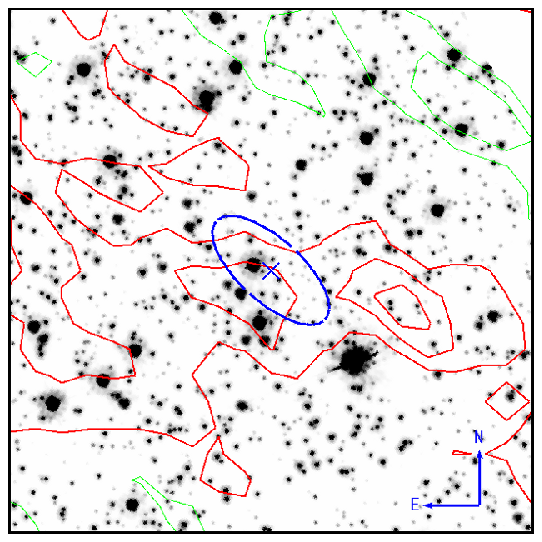}
\includegraphics[scale=1.5]{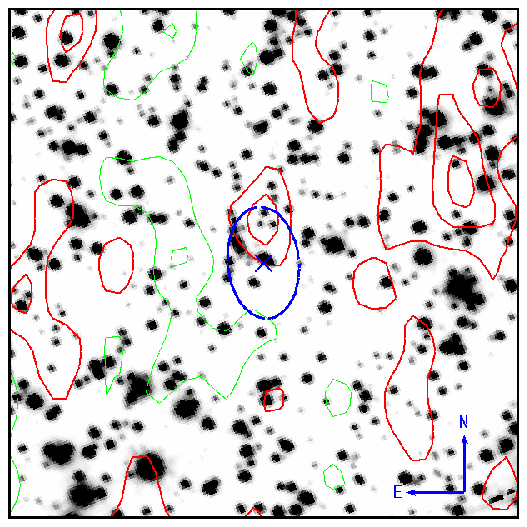}
\centering
\caption{The \hst\ R-band image overlaid with ATCA contours at frequency 6.8 GHz within the central 10$^{\prime\prime}$ $\times$ 10$^{\prime\prime}$ region of the clusters. 
The red contours are 1$\sigma$, 2$\sigma$ and 3$\sigma$ noise levels, while the green contours are -1$\sigma$ and -2$\sigma$ noise levels. 
The center of the cluster is denoted by the blue cross. 
The blue ellipse represents the 95$\%$ positional error ellipse of the globular cluster center. 
\textbf{Left} : 47 Tuc. \textbf{Right} : $\omega$ Cen.
}
\end{figure}

\begin{figure}
\includegraphics[scale=.90]{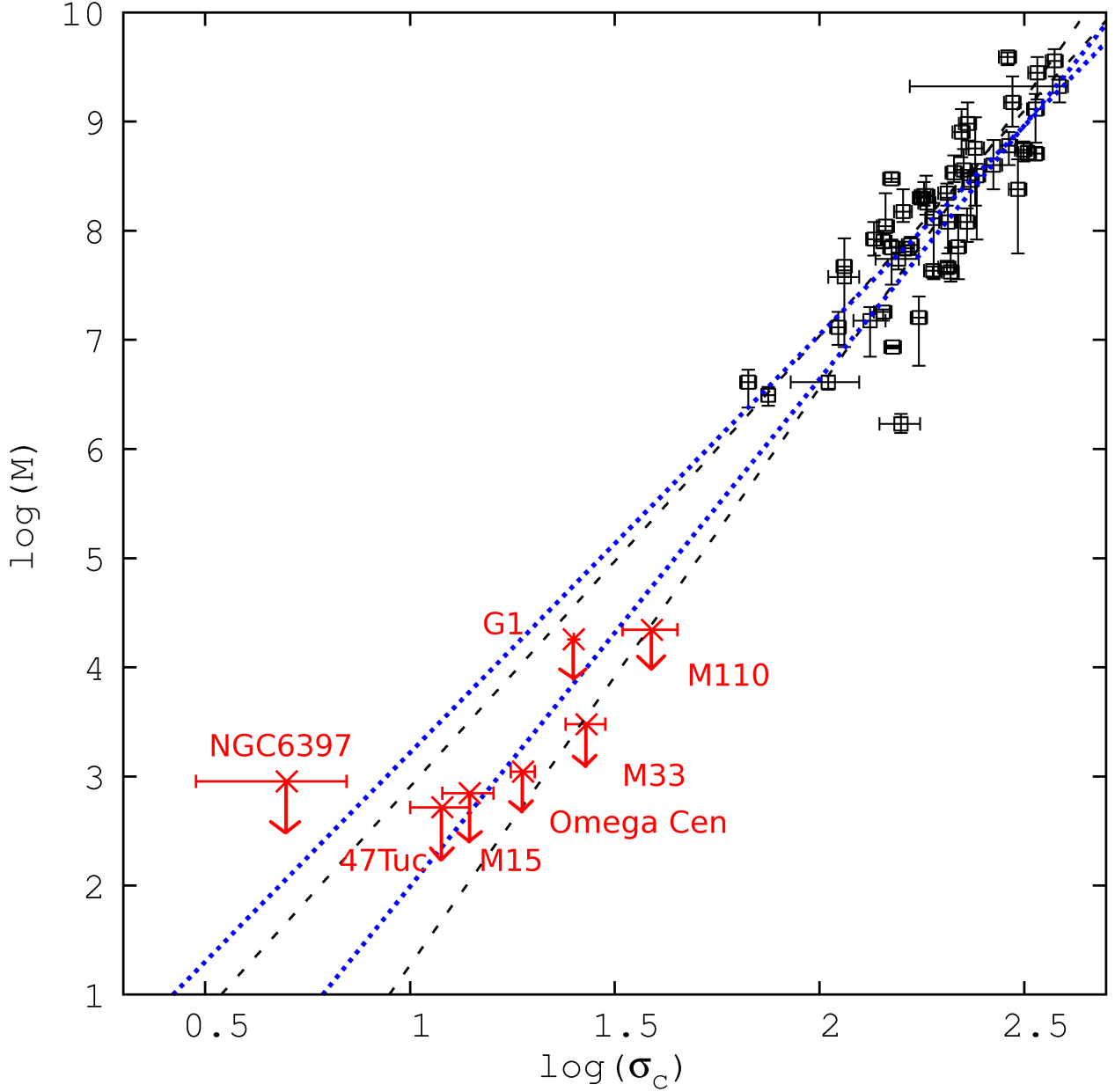}
\centering
\caption{The M-$\sigma$ relation of galaxies derived by \citet{2005SSRv..116..523F} (black-dashed line) and \citet{2009ApJ...698..198G} (blue-dotted line). The red data points represents the 3$\sigma$ upper limits on the masses of IMBHs in globular clusters, with accretion efficiency $\epsilon$ = 3$\%$. The velocity dispersion for $\omega$ Cen comes from \citet{2010ApJ...710.1063V}; for 47 Tuc comes from \citet{1995AJ....109..209G}; for M15 comes from \citep{2002AJ....124.3270G}; for NGC 6397 comes from \citet{1991A&A...250..113M}; for G1 comes from \citet{2001AJ....122..830M}; for M33 comes from \citet{2001Sci...293.1116M}; for M110 comes from \citet{2005ApJ...628..137V}.
}
\end{figure}

\end{document}